\documentclass[aps,floatfix,twocolumn,superscriptaddress,showpacs,letter]{revtex4-1}
\usepackage{graphics,graphicx,dcolumn,bm,fleqn,epic,eepic,float}
\usepackage{subfigure}
\usepackage{amssymb,amsmath,multirow,rotate,color,float}
\usepackage[latin1]{inputenc}
\usepackage[dvips]{epsfig}

\graphicspath{{}{images/}}

\def\epsilon{\varepsilon}
\def\theta{\vartheta}

\def\rho{\varrho}

\def\vec#1{\mathbf{#1}}

\newcommand{\be}{\begin{equation}}
\newcommand{\ee}{\end{equation}}

\renewcommand{\vec}[1]{\mathbf{#1}}

\begin{document}
%\title{Hydrodynamic interactions in colloidal crystals}
\title{Hydrodynamic interactions in active colloidal crystal microrheology}

\author{R.~Weeber}
\affiliation{Institute for Computational Physics, University of Stuttgart, Pfaffenwaldring 27, D-70569 Stuttgart, Germany}

\author{J.~Harting}
%\email{j.harting@tue.nl}
\affiliation{Department of Applied Physics, Eindhoven University of
  Technology, P.\,O. Box 513, 5600\,MB Eindhoven, The Netherlands}
\affiliation{Institute for Computational Physics, University of Stuttgart, Pfaffenwaldring 27, D-70569 Stuttgart, Germany}

\date{\today}

\begin{abstract}
\noindent
In dense colloids it is commonly assumed that hydrodynamic interactions do
not play a role. However, a found theoretical quantification is often
missing. We present computer simulations that are motivated by experiments
where a large colloidal particle is dragged through a colloidal crystal.
To qualify the influence of long-ranged hydrodynamics, we model the setup
by conventional Langevin dynamics simulations and by an improved scheme
with limited hydrodynamic interactions. This scheme significantly improves
our results and allows to show that hydrodynamics strongly impacts on the
development of defects, the crystal regeneration as well as on the jamming
behavior.
\end{abstract}



\maketitle

% ============================================================================

\section{Introduction}
%{\it Introduction.--}
Colloidal crystals are commonly studied complex fluids which allow to
investigate the interplay between the local micromechanics and the macroscopic
mechanical properties of the crystal. Also, they can be used as a model system
for nanoscopic systems since colloids are easier to observe and manipulate
than atoms due to their larger size and slower
dynamics~\cite{bib:prasad-2007}. Among others,
melting~\cite{bib:kesavamoorthy-1997,bib:reichhardt:2004b,bib:dullens-bechinger-2011},
defect dynamics\cite{bib:pertsinidis-2001,korda2003}, phase
behavior~\cite{Akhilesh97}, and glass formation~\cite{bib:petekidis-2004} have
been studied with colloidal crystals. They can be manipulated on the single
particle level by optical tweezers: due to electric gradient forces, particles
are trapped and moved along in the focus of a laser beam~\cite{chu1986}. For
moderate displacements from the trap center the trapping potential can be
approximated to be parabolic resulting in a force that is proportional to the
displacement. We follow recent experiments and study a system where a large
particle is dragged through a two-dimensional triangular colloidal crystal of
smaller ones. All particles are suspended in water and
slightly charged. The dynamical properties of the crystal, like stiffness and
defect formation, can be measured directly using optical
methods~\cite{bib:dullens-bechinger-2011}. 
For numerical simulations of colloidal crystals it is commonly assumed that due
to electostatic repulsion of the particles, their dense packing and the generally moderate deformation of the crystal the
system can be assumed as overdamped and long-range hydrodynamic interactions
are ignored in order to reduce the computational effort~\cite{bib:riese-2000,bib:reichhardt:2004b}. However, in case of
optical tweezer experiments the situation is less clear. The movement of the
large probe particle can deform and even locally melt the crystal. Furthermore, so
as to not influence surrounding colloidal particles by the trapping laser, the
driven particle has to be significantly larger than the surrounding ones. This
implies that it creates significant drag on the fluid. We show
that it strongly influences the healing of defects and restructuring of the
crystal. 
Similar systems have been studied numerically earlier~\cite{bib:reichhardt:2004b}. However, instead of
pushing the probe with a constant force, we model the moving
parabolic potential of the optical tweezer and use size ratios as in
experiments~\cite{bib:dullens-bechinger-2011}. 
To demonstrate the effect of long-range hydrodynamics, the system is modeled in 3D using
conventional Langevin dynamics (LD) and an improved scheme that includes
limited hydrodynamics between the probe and the surrounding
crystal~\cite{rauscher07b,bib:jens-gutsche-kremer-krueger-rauscher-weeber:2008,bib:jens-gutsche2}.
The reason for this choice is two-fold: due to its computational efficiency
conventional LD is the most commonly used method to simulate overdamped
colloids. On the other hand, the improved scheme utilizes a correction of the
velocity field as induced by the large particle. By switching between the
schemes we can investigate in a simple manner the impact of the flow
induced by the probe on the crystal itself. 

\section{Simulation technique}
%{\it Simulation technique.--}
\label{sec:simulationMethod}
When a probe is dragged through a colloidal crystal, the dynamics is
influenced by several factors: strong forces due to the tweezer and the
electrostatic potentials between the particles exist. These forces dominate the
dynamics in front of the probe. 
Furthermore, especially the particles behind it feel a
hydrodynamics-mediated drag force in the direction of motion. This drag also
exists in front of the probe, but there the system is more jammed. Behind the
probe, diffusion and electrostatic repulsion among particles control
the ``healing'' of the crystal. Our simulations utilize a
modified LD method, which gives more accurate results for
the drag force on the probe and to some extent models the drag exerted behind
the driven particle~\cite{rauscher07b}. To demonstrate the importance of long-ranged
hydrodynamic effects in colloidal crystals, we compare our results to
conventional LD.
For overdamped systems computing time can be saved by
not explicitly calculating the flow field. Instead, only the most relevant
effects of the fluid on suspended particles are simulated by additional
forces in a molecular dynamics (MD) algorithm: the Stokes friction and thermal
fluctuations. The motion of the $i$'th particle is described by the Langevin
equation
\begin{equation}
 m \ddot{\vec{x}}_i = \gamma \dot{\vec{x}}_i +\vec{F}_i^{\rm rnd}+\vec{F}_i^0,
\end{equation}
where $\gamma$ is the friction coefficient, $\vec{F}_i^{\rm rnd}$ is a random
force describing Brownian motion, and $\vec{F}_i^0$ represents forces between
particles (e.g. a Coulomb force) and external forces (e.g. gravity). 
In case of fully overdamped dynamics the inertia term can be dropped reducing
the method to simplified Brownian Dynamics. However, due to the movement of
the probe particle inside the optical trap inertia should not be ignored.
According to the fluctuation-dissipation-theorem, the amplitude of the random
force is connected with the friction coefficient of the Stokes force. It is
assumed that $\vec{F}_i^{\rm rnd}$ is Gaussian-distributed with zero mean and
uncorrelated in time. This is a strong approximation: if one particle starts
moving, it drags some of the surrounding fluid with it. This in turn drags
along secondary particles in its vicinity.
The mean square deviation of $\vec{F}_i^{\rm rnd}$ is given by
% the fluctuation-dissipation theorem as
\begin{equation}
  \langle| \vec{F}_i^{\rm rnd}|^2 \rangle = 12\,\pi\,\eta\,R\,k_BT.
\end{equation}
LD is popular to simulate suspensions due to its simplicity and low amount
of computation required. However, it completely lacks the
hydrodynamic interactions between particles. In the system we consider, these
interactions are, however, important: the large particle has a volume of about
125 times that of the small ones and also moves at a substantially higher
velocity. Thus, it strongly influences the flow around it.  Furthermore, the
surrounding particles are dragged due to the motion of the probe and the
flow advects crystal particles to the sides at the front of the dragged
particle and back into the empty region behind it. The smaller and slower
particles do not have as large impact on the flow. Therefore, it is
possible to improve the LD scheme by including the flow field created by the
large particle~\cite{rauscher07b}: the probe still feels the Stokes
friction imposed by a resting fluid with viscosity $\eta$. Small particles with velocity $\vec{v}$ and radius $R_{\rm
small}$, however, feel a force 
%\be
%\vec{F}=6\pi\eta R_{\rm small} (\vec{v} -\vec{v}_f)
%\ee
$\vec{F}=6\pi\eta R_{\rm small} (\vec{v} -\vec{v}_f)$
due to a moving fluid as caused by the motion of the probe. Here,
\begin{equation}
\label{method_sphereflow}
\vec{v}_f(\vec{r}) =\\
\frac{3\,R}{4\,r}\,\left[
\left(1+\frac{R^2}{3\,r^2}\right)\,
\vec{u} + 
\left(1-\frac{R^2}{r^2}\right)\,\vec{\hat{r}}\,(\vec{\hat{r}}\cdot\vec{u})
\right]
\end{equation} 
is the fluid velocity at the position of the small
particle~\cite{bib:landau-lifshitz-fluid-mechanics}, where $\vec{\hat{r}}$ is
the unit vector in the direction from the large to the small
particle, $r$ is their distance, and $R$ and $\vec{u}$ are the large particle's
radius and velocity. As shown in this letter, this scheme significantly
improves the results for suspensions, through which a colloidal particle is
dragged by an optical tweezer. The streamlines on the upwind side of the probe
bend around it. Thereby, they also move obstacles out of the way of the probe
lowering the drag exerted on it. Also, the flow field created by the large
particle drags along some of the surrounding crystal causing the movement of an
effectively much larger object.
In particular behind the probe
the drag is increased, since there the crystal is less
jammed~\cite{bib:jens-gutsche-kremer-krueger-rauscher-weeber:2008,bib:jens-gutsche2}.
The particles forming the crystal have a radius of 1.6$\mu$m and are kept
on the bottom of the container by gravity. The surface area of
the bottom of the container that has to be covered to form a stable crystal depends on
the interaction potential. We consider weakly charged particles and
area densities of $\sim 74 \%$ resulting in a lattice spacing of $a=3.5 \mu$m.  
%For the
%initial condition, a circular area with a radius slightly larger than that
%of the large particle has to be cleared from small particles because
%otherwise the strong forces caused by overlapping particles would cause
%the MD algorithm to become unstable.  
%
The crystal is generated by explicitly calculating the particle positions,
where the stacking distance between horizontal rows is $a \sin
60^{\circ} =\frac{\sqrt{a}}{2} \approx 0.866a$ and the offset
in consecutive rows is $a \cos 60^{\circ} =\frac{a}{2}$.
A screened Coulomb potential
\be
V = \frac{1}{r}\exp\left(-\kappa (r-R_1-R_2)\right),
\ee
%$V = r^{-1}\exp\left(-\kappa (r-R_1-R_2)\right)$,
% \sim or = ?
is used to model the interactions, where $\kappa$
is the inverse screening length and $R_1$ and $R_2$ are the particle radii.
The potential at the surface of the colloids is 10mV and the screening
length is varied between 20 and 50nm. Identical $\kappa$ are chosen for
the interactions between small particles and between a small and the
large particle, because screening is controlled by the ion concentration
in the solvent rather than by a property of the colloidal particles.  The
dragged particle has a radius $R=7.75\mu$m and the potential of the
optical tweezer
\be
\label{eq_trapPotential}
 V_{\rm trap} =C \cdot (\vec{x}-\vec{x}_t(t))^2, 
\ee 
%$V_{\rm trap} =C \cdot (\vec{x}-\vec{x}_t(t))^2$ 
is assumed to be harmonic, where $\vec{x}$ and $\vec{x}_t$ are the positions of the particle,
and trap center and $t$ is time. The force constant
is chosen as $C = 3.2 \cdot 10^{-7}$Jm$^{-2}$, which is a reasonable value
for experiments: a particle with a distance of 1$\mu$m from the trap
center feels a potential of $\approx 40 k_B T$. The tweezer moves through the crystal
with a constant velocity between $0.5\mu$m/s and $8\mu$m/s.
A long system is required since it takes up to 240s to reach the steady state
-- especially for simulations at low velocities including hydrodynamic
corrections.  I.e., for a velocity of 4$\mu$m/s, the probe passes approx.
1000$\mu$m, before the trapping force equals the friction on the probe. If the
system is too narrow in the direction perpendicular to the drag, it gets jammed
and the steady state is not reached before the probe arrives at the end of the
channel. Furthermore, long-ranged electrostatic and hydrodynamic interactions can
cause finite size effects. Due to strong fluctuations of the measured probe
position, averaging over 60 to 100s is needed for drag force measurements
rendering our simulations computationally demanding. We use a crystal of
approx. 570 $\times$ 165 particles
%, with a lattice constant of $a=3.5\mu$m,
resulting in a system size of 2000$\mu$m $\times$ 500$\mu$m. The system height and timestep are 30$\mu$m and $\delta t=98.4\mu$s. To allow for such a large time step -- and therefore
reducing computation time -- we rescale some physical units: the suspension is
simulated at a lower temperature and viscosity. To preserve the P\'eclet
number, i.e., the relative importance of dynamics due to potentials and
diffusion, all forces are scaled down by the same factor (here 1/11933) as described
in~\cite{bib:jens-hecht-ihle-herrmann:2005}. 
%Here, the scaling factor is 1/11933.

\begin{figure}[thb]
%\begin{tabular}{p{0.1\linewidth}p{0.4\linewidth}p{0.1\linewidth}p{0.4\linewidth}}
\begin{tabular}{llll}
a)& &b)&\vspace*{-0.3cm}\\
%&\includegraphics[width=0.33\linewidth]{density-maps/hydro-00_grad_8E-06_kappa_2E+07_offset_0}&
&\includegraphics[width=0.43\linewidth]{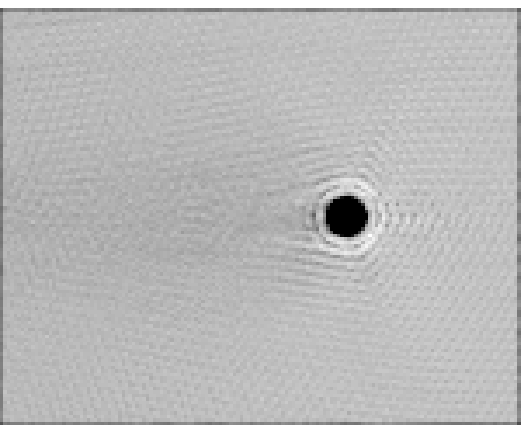}&
%b)&
%&\includegraphics[width=0.33\linewidth]{density-maps/hydro-00_grad_8E-06_kappa_5E+07_offset_0}
&\includegraphics[width=0.43\linewidth]{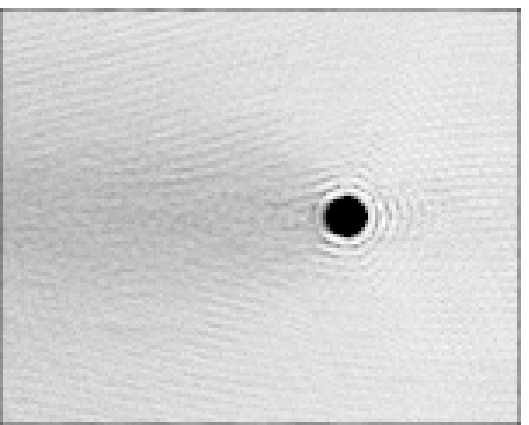}
\\
c)& &d)&\vspace*{-0.3cm}\\
%&\includegraphics[width=0.33\linewidth]{density-maps/nohydro-00_grad_8E-06_kappa_2E+07_offset_0}&
&\includegraphics[width=0.43\linewidth]{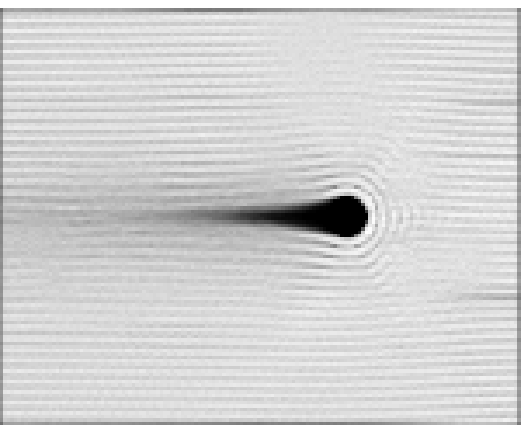}&
%d)&
%&\includegraphics[width=0.33\linewidth]{density-maps/nohydro-00_grad_8E-06_kappa_5E+07_offset_0}
&\includegraphics[width=0.43\linewidth]{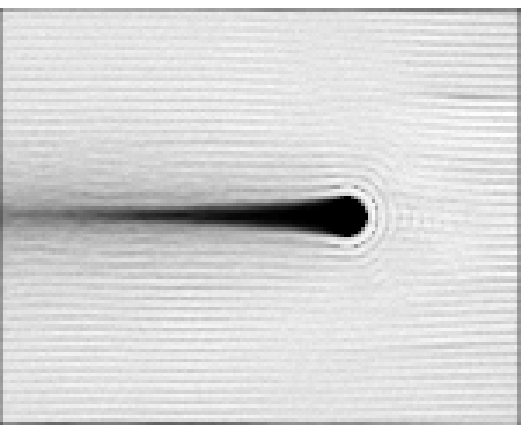}
\end{tabular}
\caption{\label{fig:densitymap-no-hydro}
Density map for $\vec{u}=$8$\mu$m/s and $1/\kappa=$50nm (left) and 20nm (right)
with (upper row) and without (bottom row) hydrodynamic correction.  The crystal
melts in front of the probe and small particles rearrange into a circular
structure. When neglecting the flow field of the probe, the lack of drag and
the advection of the small particles around it cause a large depletion zone to
form. With correction, this zone is substantially reduced. Healing is more
efficient for larger $1/\kappa$ since particles are pushed into the empty
region by electrostatic repulsion.
}
\end{figure}

\section{Results}
%{\it Results.--}
\label{sec:hydro}
The volume of the probe particle is by a factor of 125 larger 
than that of the small ones. When it moves, it creates a
significant flow around it inducing two consequences: crystal
particles at the front are advected to the sides and around the probe. They
then fill the depletion region and thereby weaken it. The drag
force felt by the probe is reduced since the obstacles are moved to the
sides without getting in contact with the probe. Furthermore, the
crystal particles around the probe are dragged along with its motion leading to an
effectively much larger object that is dragged through the crystal. Thus,
more reorganisation has to take place leading to an increased drag. 
The influence of this flow field can be investigated quantitatively by comparing
simulations using LD with the flow field correction
as described above and conventional LD which does not
include any hydrodynamics. Other hydrodynamic
effects -- like the influence of the surface on which the crystal lies -- are
not taken into account here. We first turn to the surrounding of
the probe and then elaborate on the quantitative effect of
the flow field on the drag force exerted on the probe.

\begin{figure}[htb]
\begin{tabular}{ll}
a)&\vspace*{-0.3cm}\\
&\includegraphics[width=0.28\linewidth]{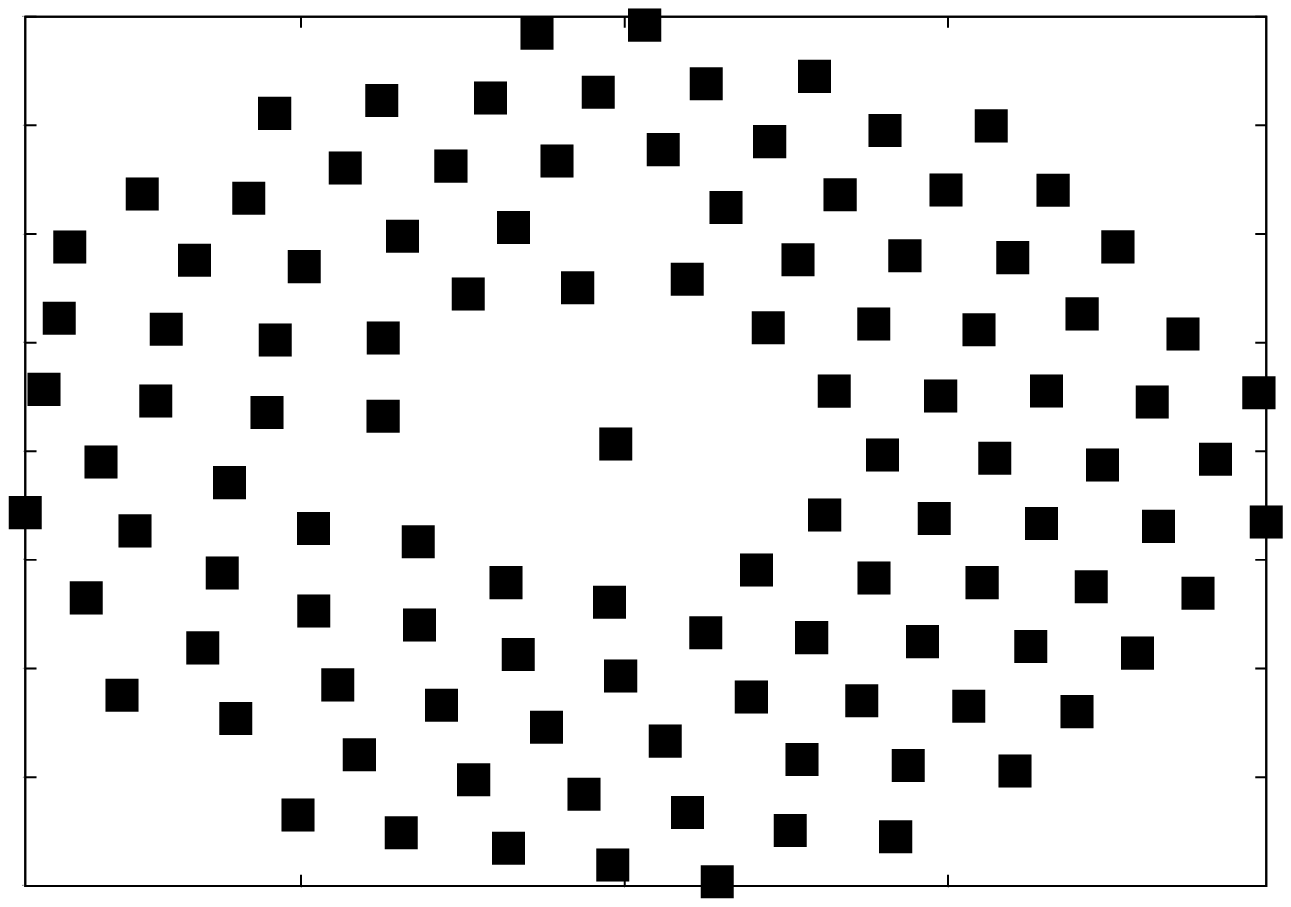}
\includegraphics[width=0.28\linewidth]{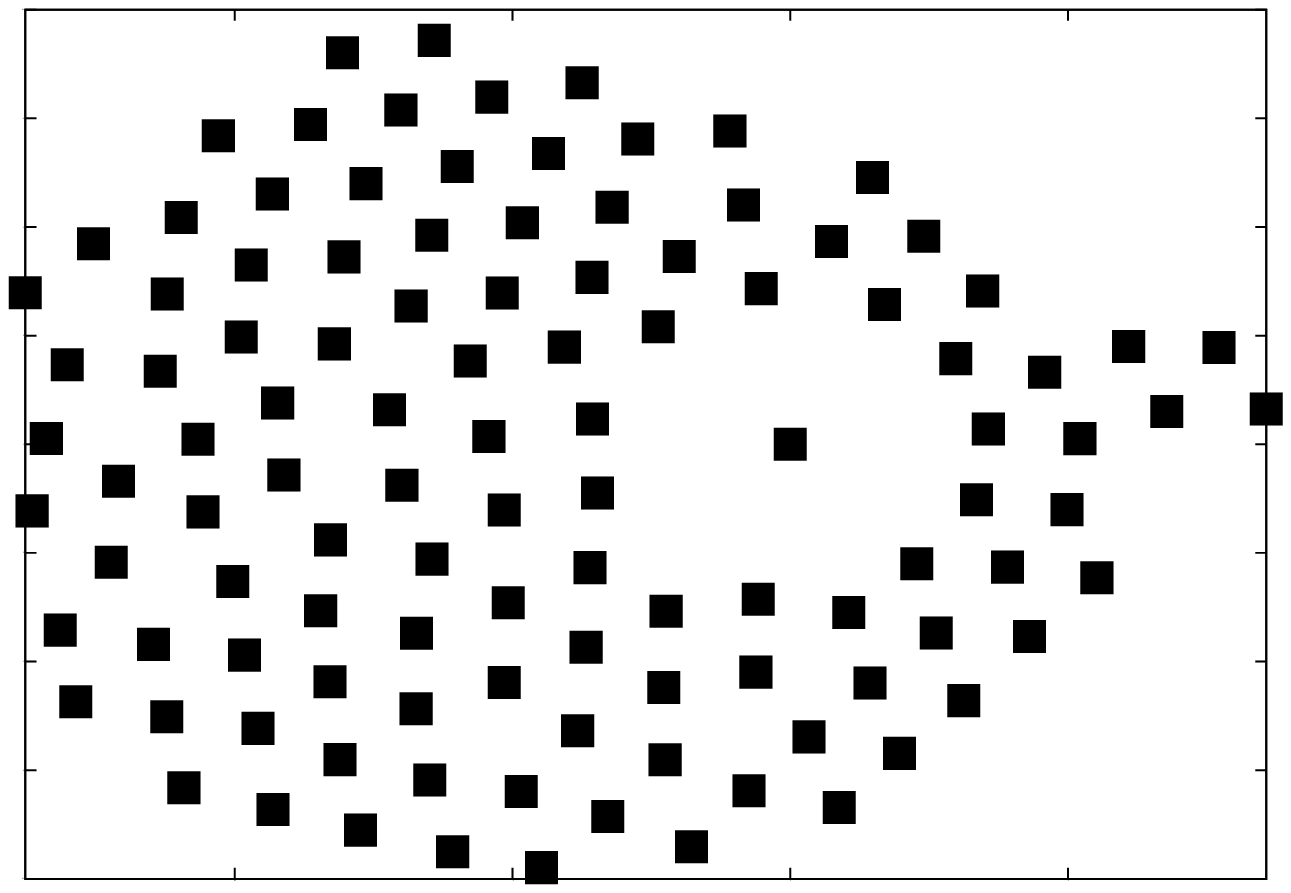}
\includegraphics[width=0.28\linewidth]{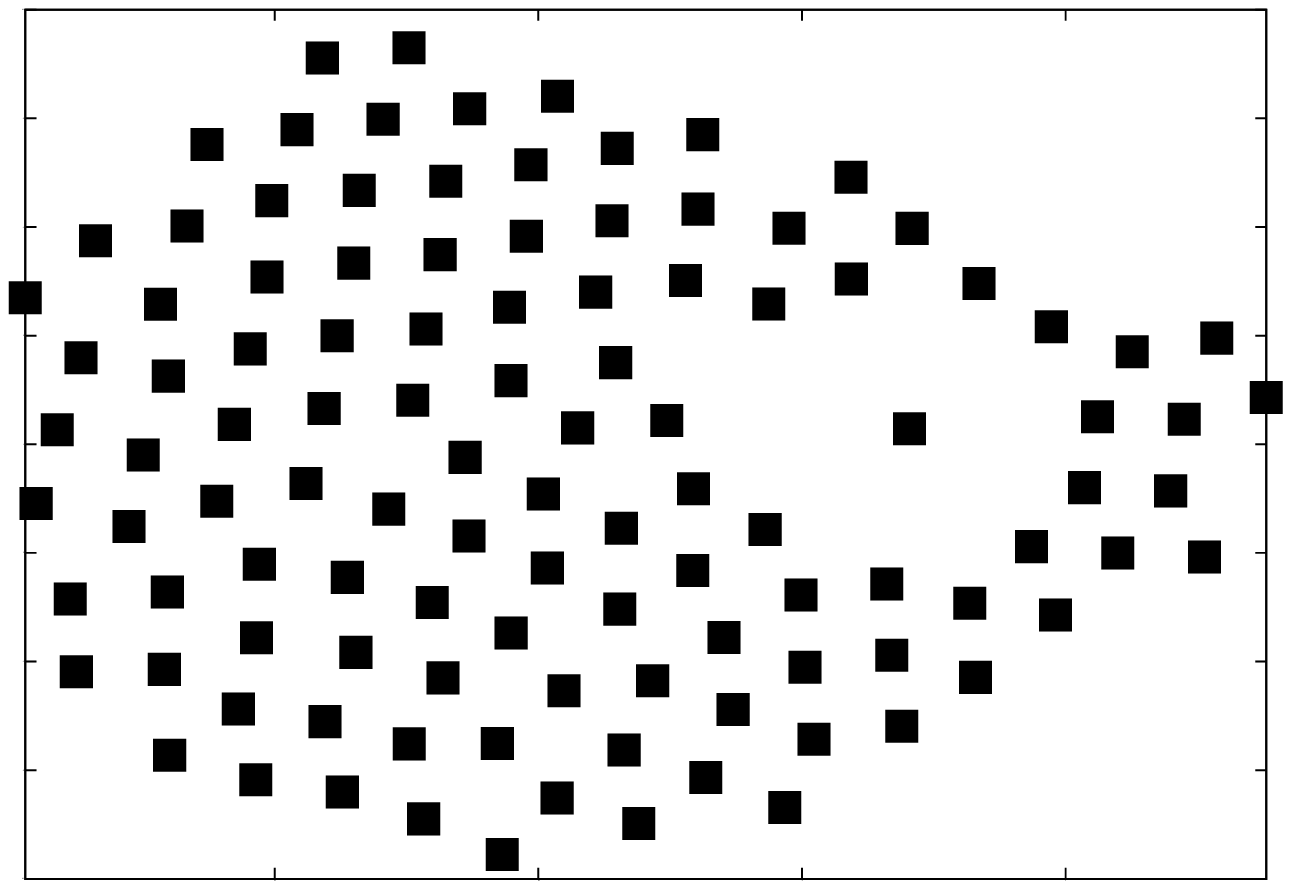}\\
b)&\vspace*{-0.3cm}\\
&\includegraphics[width=0.28\linewidth]{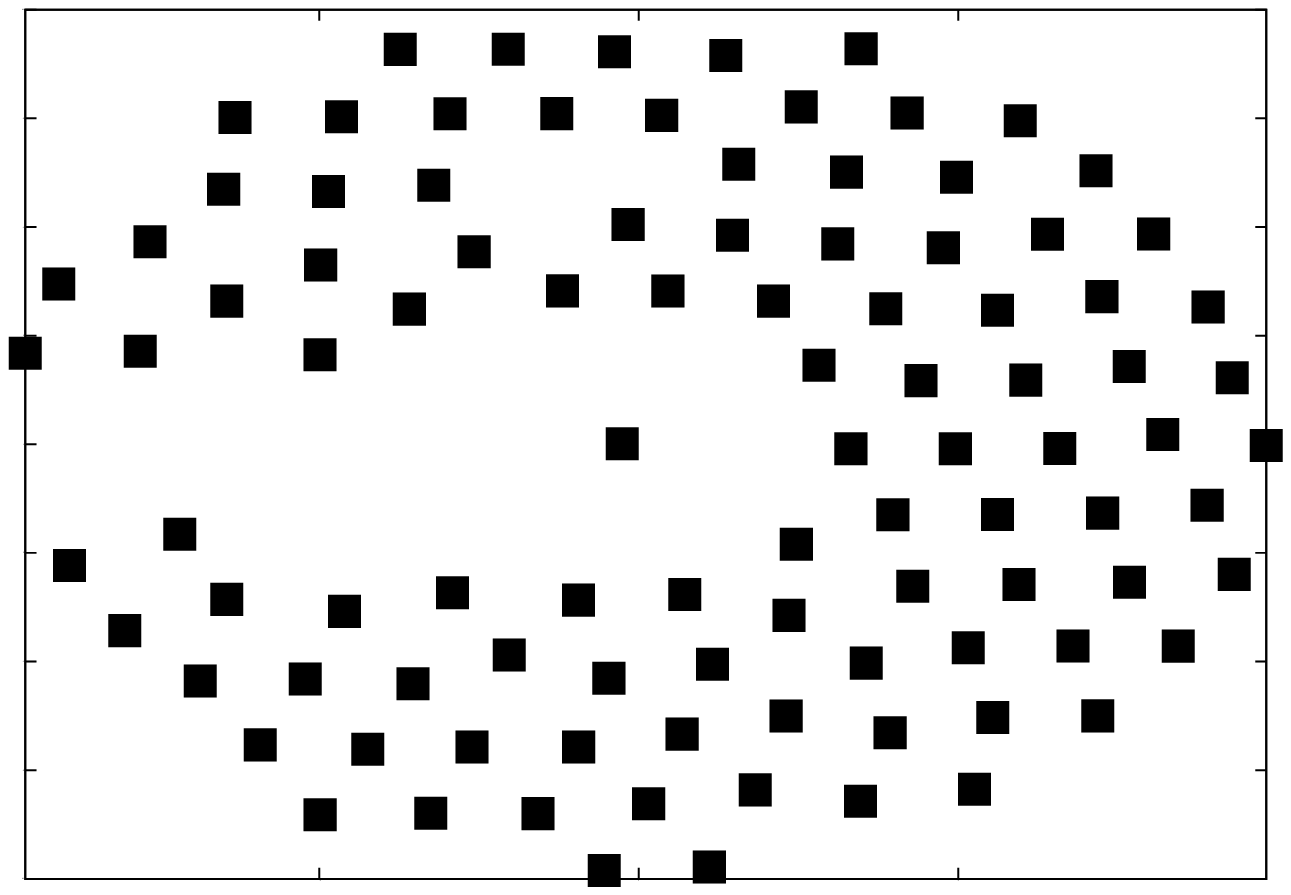}
\includegraphics[width=0.28\linewidth]{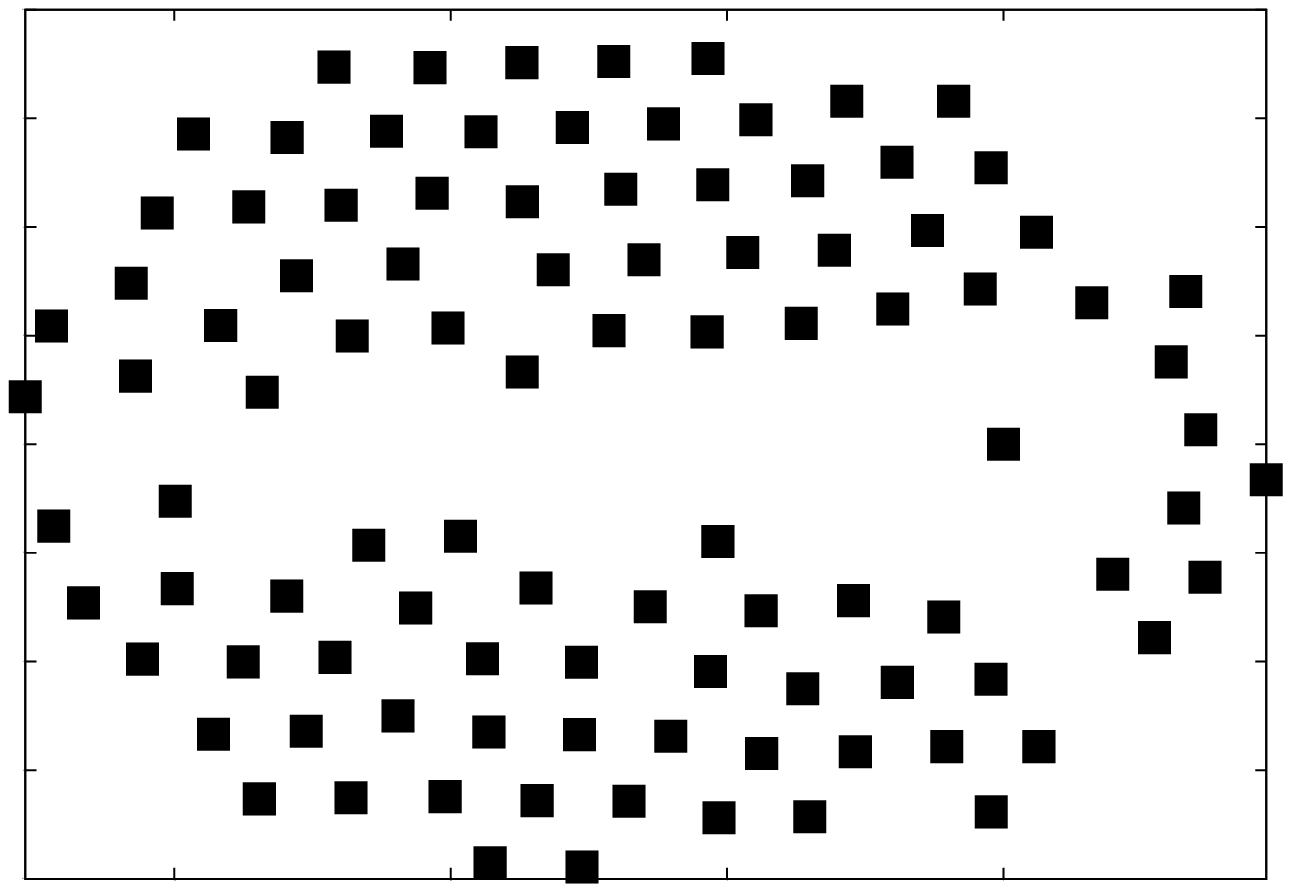}
\includegraphics[width=0.28\linewidth]{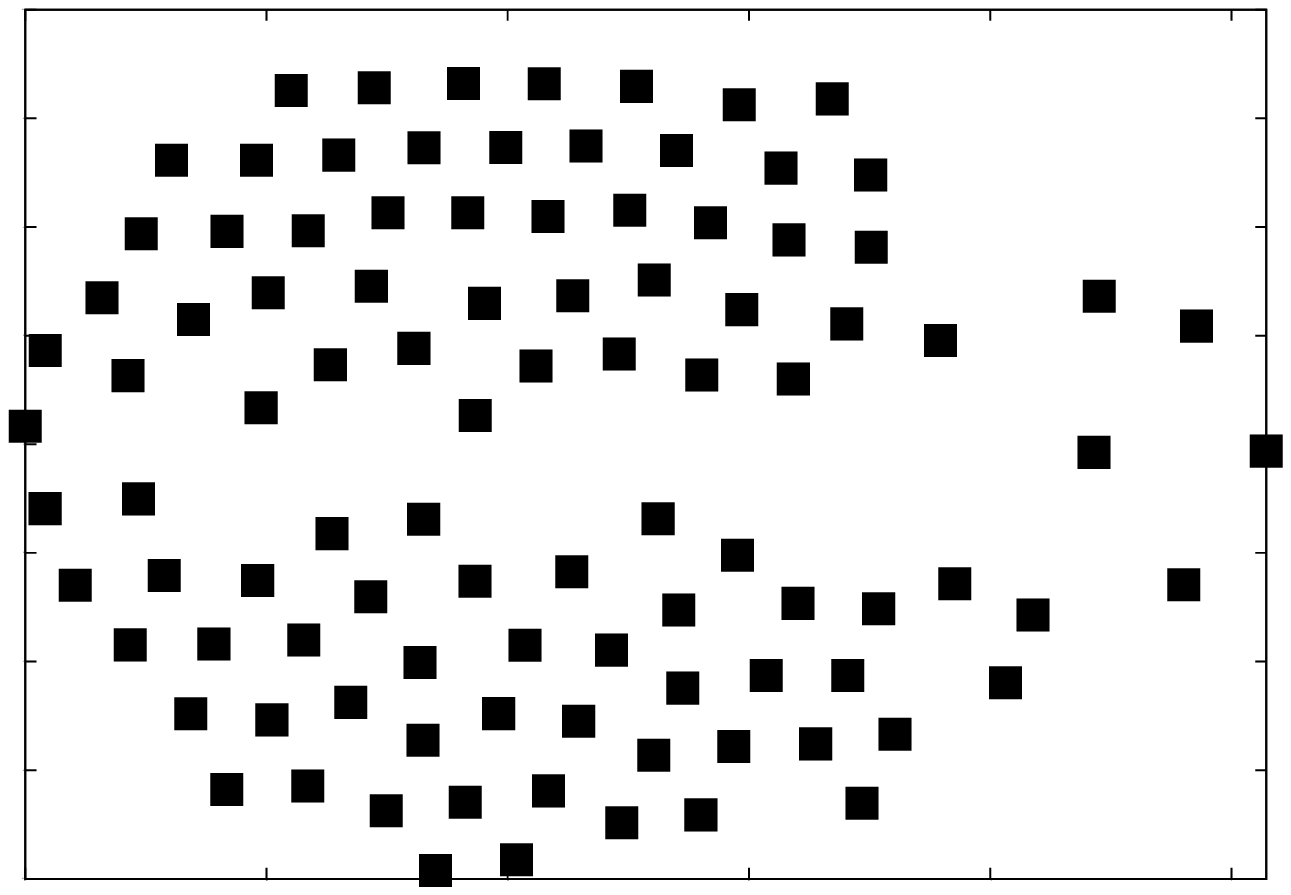}\\
\end{tabular}
\caption{\label{fig:trace}
Traced particle positions around the probe with (a) and
without (b) the flow field of the probe taken into account. The probe moves at
5$\mu$m/s. In the first time step, particles that are within a distance of
20$\mu$m of the probe's center are marked (left). These are
observed after 3s (middle) and 5s (right). If the probe's flow
field is considered, the probe is not moving individually through the crystal,
but is accompanied by an entire cloud of particles. This effect is only weakly
present in conventional LD simulations.
}
\end{figure}

In Fig.~\ref{fig:densitymap-no-hydro}, we compare density maps for the case
with and without the probe's flow field considered. To obtain such a map,
about 1000 snapshots of the crystal are taken while the probe is driven through
it and moved such that the driven particle coincides in all of them. We divide
the system into 150 $\times$ 120 bins and calculate the probability for each
bin to be occupied. It can be seen that for conventional LD, a
large depletion zone forms behind the probe due to the lack of slip
stream, i.e. particles do not diffuse into the depletion region. We find
that this region is the more pronounced, the higher the velocity
gets, because then, the probe clears more space in a given
time, whereas the reconstruction process behind the probe is not
directly influenced by the velocity. It is, however,
controlled by three factors: diffusion, electrostatic repulsion between the
crystal particles and the drag created by the driven particle. Only
the latter is directly affected by $\vec{u}$.
In the case with the flow field modeled, however, this depletion zone is
substantially reduced.
The circular structure created by the melting of the crystal close to the
probe's surface is visible in both maps. At long-ranged potentials and
high velocities, layers of circularly arranged particles can be
seen. With hydrodynamic correction, however,
the circular structure extents further back because the flow
advects crystal particles around the probe and into the
depletion zone behind it.

In Fig.~\ref{fig:trace}, we consider individual particles close to the probe's
surface while it moves with a velocity of 5$\mu$m/s.
Thereby, we study how the dynamics of the crystal particles changes
due to the hydrodynamic correction. In the first time step, all particles that
lie within 20$\mu$m of the probe's center are tagged. We then follow these
particles in time. In Fig.~\ref{fig:trace}, the particles are
shown after 3s and 5s for both, a simulation with and without hydrodynamic
correction. It can be seen that when the flow field evolving
around the probe is considered, the probe does not move through the crystal
individually. Rather, it is accompanied by an entire cloud of smaller
particles. In a conventional LD simulation, on the other hand, the surrounding
particles drop behind the moving probe.

\begin{figure}[hbt]
\includegraphics[width=0.9\linewidth]{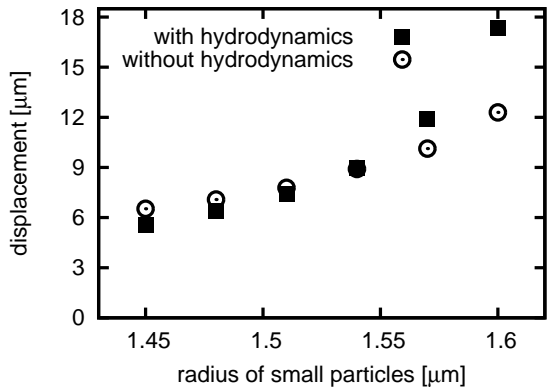}
\caption{
\label{fig:drag-radius}
Displacement of the probe versus the small particles' radius
with and without hydrodynamics considered. All simulations were
performed at a drag velocity of 4$\mu$m/s and $1/\kappa=$40nm. It
can be seen that for smaller radii, the hydrodynamic interactions lower the drag
force acting on the probe due to the advection of crystal particles around it.
For larger radii, the system becomes jammed and hydrodynamics results in an
additional friction due to the increased level of rearrangement
in the surrounding crystal.
}
\end{figure}

The interplay between a lower drag force due to the advection of the
surrounding crystal particles and an increase due to part of
the crystal moving along is studied by comparing the forces at different
radii of the small particles. In Fig.~\ref{fig:drag-radius}, the displacement
of the probe is shown versus radii
of the small particles between 1.45$\mu$m and 1.6$\mu$m, both, with and without
hydrodynamic correction. At a radius of about
1.54$\mu$m, the same drag is obtained for both cases. Below this point,
the force is lower if hydrodynamic corrections are applied because of
the advection of crystal particles around the probe. For larger radii, on the
other hand, the crystal is more jammed and due to the dragging along of
surrounding small particles, a higher drag force is exerted on the probe.
%
%Finally, we study quantitatively the reduction in drag force acting on
%the probe due to the surrounding particles being dragged along and advected
%around the probe. In Fig.~\ref{fig:drag-hydro-comparison},
%force-velocity curves are compared for improved and conventional LD
%and for $1/\kappa=$20 and 50nm. 
%It can be seen that the static
%friction effect is weaker if the hydrodynamic correction is not taken into
%account since the slope of the curve does not change between
%low and high velocities.
 
\begin{figure}
\includegraphics[width=0.9\linewidth]{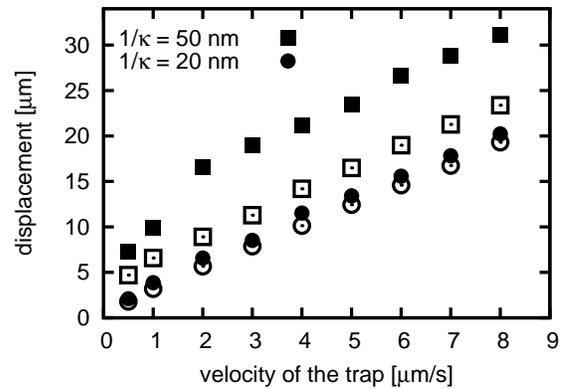}
\caption{
\label{fig:drag-hydro-comparison}
Comparison of probe displacement in the trapping potential depending on the
drag velocity from simulations with (full symbols) and without (open symbols)
hydrodynamic correction and $1/\kappa=$20 and 50nm. In all cases, the
displacements for the simulations with hydrodynamic correction are higher than
for the case with conventional LD.  This is because, as can be seen from
Fig.\,\ref{fig:trace}, due to the hydrodynamic corrections, not only the probe
but also a large number of close by crystal particles are moving. Therefore,
much more rearranging has to take place to allow them to pass. 
}
\end{figure}

Finally, we study quantitatively the reduction in drag force acting on
the probe due to the surrounding particles being dragged along and advected
around the probe. Data has been obtained for $1/\kappa=$20, 30 and 50nm as well
as drag velocities between 0.1 and 10$\mu m/s$. In
Fig.~\ref{fig:drag-hydro-comparison},
force-velocity curves are compared for improved and conventional LD.
For a clear presentation, data for $1/\kappa=$30nm and some intermediate
values are omitted, but consistently fit into the presented range of results. 
It can be seen that the static friction effect is weaker if the hydrodynamic
correction is not taken into account. This can be understood from
Fig.\,\ref{fig:trace}: due to the hydrodynamic corrections, not only the probe
but also a large number of close by crystal particles are moving.  This leads
to a much stronger rearrangement to be necessary in order to allow the
particles to pass.

The experiments in~\cite{bib:dullens-bechinger-2011} report a plateau of the
force versus drag velocity curves for very small drag velocities $\le$ 0.2$\mu$m/s and the
authors suggest a finite yield stress as an explanation of this finding. Below
a critical drag velocity, the inserted energy is mainly dissipated due to a
local distortion of the colloidal crystal. In our simulations we are not able
to reproduce this phenomenon within the limits of computationally feasible
minimum drag velocities and statistical averaging. If a plateau would occur in
our data for velocities below 0.5$\mu$m/s, it would be hidden by the error
bars of our measurements. However, our simulations do confirm the experimental
finding of a power law behavior of the force versus drag velocity measurements.
The exponents strongly depend on the inverse screening length $1/\kappa$. With
hydrodynamic corrections we find 0.77 for $1/\kappa$=50nm, 0.54 for
$1/\kappa$=20nm, where the latter is close to the value of 0.51 as reported
in~\cite{bib:dullens-bechinger-2011}. Without taking the flow field of the
probe particle into account we obtain 0.86 and 0.62, respectively.

\section{Conclusion}
%{\it Conclusion.--}
We conclude that in drag experiments as discussed in this letter, the
assumption of the system to be overdamped is not valid and hydrodynamic
interactions must not be neglected. By applying a modified LD scheme which
includes the influence of the flow field around the driven probe, we
demonstrated that the dynamics of the probe and the surrounding crystal are
strongly influenced by the hydrodynamic interactions: the crystal particles
around the probe are dragged along by its slip stream and advected around it.
This significantly reduces the size of the depletion zone behind the probe.
The drag force acting on the probe is increased or decreased depending on the
density of the crystal. 

Although the flow field generated by the moving probe is the most
important hydrodynamic effect, efforts should be undertaken to perform
simulations with full hydrodynamics using for example a combined method
utilizing MD for the particles and a mesoscopic solver for the
hydrodynamics~\cite{bib:jens-hecht-ihle-herrmann:2005,bib:jens-jansen-2011,Ladd01}.
This is, however, not easy, because the time step in the molecular
dynamics simulation has to be small to properly model the steep Yukawa
potentials.  Therefore, the flow field would either have to be updated
very often -- which requires huge computational resources, or the
simulation would have to use different time steps for MD and hydrodynamics
-- which might introduce unwanted artefacts. Furthermore, in particular for
very low drag velocities, thermal fluctuations cause the necessity of
averaging over very long simulation times in order to obtain reliable
data. Another obstacle is the very different sizes of large and small
particles which would force one to use a very small lattice spacing, if
the small particles are still to be resolved.

\begin{acknowledgments}
%{\it Acknowledgments.--} 
We thank M.~Rauscher, R.~Dullens, and C.~Bechinger for fruitful
discussions and FOM (IPP IPoGII) and NWO/STW (VIDI grant 10787 of
JH) for financial support. Computations were performed at the J{\"u}lich
Supercomputing Centre and the Scientific Supercomputing Centre Karlsruhe.
\end{acknowledgments}

%\bibliographystyle{abbrv-unsrt-notitle}
%\bibliography{main_updated,jens-pub}

\begin{thebibliography}{10}

\bibitem{bib:prasad-2007}
V.~Prasad, D.~Semwogerere, and E.~R. Weeks.
\newblock {\em J. Phys. Condens. Matter}, 19:113102, 2007.

\bibitem{bib:kesavamoorthy-1997}
R.~Kesavamoorthy and C.~B. Rao.
\newblock {\em Bulletin of Materials Science}, 20:565, 1997.

\bibitem{bib:reichhardt:2004b}
C.~Reichhardt and C.~J. Olson Reichhardt.
\newblock {\em Phys. Rev. Lett.}, 92:108301, 2004.

\bibitem{bib:dullens-bechinger-2011}
R.~P.~A.~Dullens and C.~Bechinger.
\newblock {\em Phys. Rev. Lett.}, 107:138301, 2011.

\bibitem{bib:pertsinidis-2001}
A.~Pertsinidis and X.~S. Ling.
\newblock {\em Phys. Rev. Lett.}, 87:098303, 2001.

\bibitem{korda2003}
P.~T. Korda and D.~G. Grier.
\newblock {\em J. Chem. Phys.}, 114:7570, 2001.

\bibitem{Akhilesh97}
A.~K. Arora and B.~V.~R. Tata.
\newblock {\em Adv. in Col. Int. Sci.}, 78:49, 1998.

\bibitem{bib:petekidis-2004}
G.~Petekidis, D.~Vlassopoulos, and P.~N. Pusey.
\newblock {\em J. Phys. Condens. Matter}, 16:S3955, 2004.

\bibitem{chu1986}
A.~Ashkin, J.~M. Dziedzic, J.~E. Bjorkholm, and S.~Chu.
\newblock {\em Opt. Lett.}, 11:288, 1986.

\bibitem{bib:riese-2000}
D.~O. Riese, G.~H. Wegdam, W.~L. Vos, R.~Sprik, D.~Fenistein, J.~H.~H.
  Bongaerts, and G.~Gr\"ubel.
\newblock {\em Phys. Rev. Lett.}, 85:5460, 2000.

\bibitem{rauscher07b}
M.~Rauscher, M.~Kr{\"u}ger, A.~Dominguez, and F.~Penna.
\newblock {\em J. Chem. Phys.}, 127:244906, 2007.

\bibitem{bib:jens-gutsche-kremer-krueger-rauscher-weeber:2008}
C.~Gutsche, F.~Kremer, M.~Kr\"uger, M.~Rauscher, J.~Harting, and R.~Weeber.
\newblock {\em J. Chem. Phys.}, 129:084902, 2008.

\bibitem{bib:jens-gutsche2}
C.~Gutsche, M.~M.~Elmahdy, K.~Kegler, I.~Semenov, T.~Stangner, O.~Otto, O.~Uebersch\"ar, U.~F.~Keyser, M.~Kr\"uger, M.~Rauscher, R.~Weeber, J.~Harting, Y.~W.~Kim, V.~Lobaskin, R.~R.~Netz, K.~Kremer.
\newblock {\em J. Phys. Condens. Matter}, 23:184114, 2011.

\bibitem{bib:landau-lifshitz-fluid-mechanics}
L.~D. Landau and E.~M. Lifshitz.
\newblock {\em Fluid mechanics}.
\newblock Pergamon Press, 1959.

\bibitem{bib:jens-hecht-ihle-herrmann:2005}
M.~Hecht, J.~Harting, T.~Ihle, and H.~J. Herrmann.
\newblock {\em Phys. Rev. E}, 72:011408, 2005.

\bibitem{bib:jens-jansen-2011}
F.~Jansen and J.~Harting.
\newblock {\em Phys. Rev. E}, 83:046707, 2011.

\bibitem{Ladd01}
A.~J.~C. Ladd and R.~Verberg.
\newblock {\em J. Stat. Phys.}, 104:1191, 2001.

\end{thebibliography}

\end{document}